\documentclass[epj,nopacs]{svjour}
\usepackage{amsmath}
\usepackage{amssymb}
\usepackage{latexsym}
\usepackage[dvips]{graphicx}
\usepackage[english]{babel}

\def\bge{\begin{equation}}
\def\ene{\end{equation}}
\def\bgea{\begin{eqnarray}}
\def\enea{\end{eqnarray}}

\newcommand{\hii}{H\textsc{II}}
\newcommand{\hb}{\mathrm{H}\beta}

\def\bge{\begin{equation}}
\def\ene{\end{equation}}
\def\bgea{\begin{eqnarray}}
\def\enea{\end{eqnarray}}

\def\ls{\raise 1.5pt\hbox{$\,<\;$}\kern -10.5pt\lower3.5pt
          \hbox{$\sim$}\kern 1.5pt} %%% less or similar
\def\gs{\raise 1.5pt\hbox{$\,>\,$}\kern -9.5pt\lower3.5pt
          \hbox{$\sim$}\kern 1.5pt} %%% greater or similar

\usepackage{color}

\usepackage{ulem} % use: \sout{Striked out text}

\begin{document}
\sloppy
\title{The energy conditions and model selection \\ in the local Universe}
\author{Namit Chandak$^1$, Fulvio Melia$^2$\thanks{John Woodruff Simpson Fellow.} and
Jun-Jie Wei$^{3,4}$}
\institute{$^1$Department of Physics, The University of Arizona, Tucson AZ 85721,\\
$^2$Department of Physics, the Applied Math Program, and Department of Astronomy, \\
\null\hskip 15pt The University of Arizona, Tucson, AZ 85721, \email{fmelia@email.arizona.edu}\\
$^3$Purple Mountain Observatory, Chinese Academy of Sciences, Nanjing 210023, China \\
$^4$School of Astronomy and Space Sciences, University of Science and
Technology of China, \\
\null\hskip 15pt Hefei 230026, China
}

\authorrunning{Chandak, Melia and Wei}
\titlerunning{Energy conditions in the local Universe}

\date{\today}

\abstract{The four principal energy conditions (ECs) in general relativity prohibit negative
energies, repulsive gravity and superluminal energy flows. One must invoke exotic 
matter to violate any one of these, yet $\Lambda$CDM does so quite prominently 
during inflation and in the epoch of dark energy dominance. In this paper, we 
carry out model selection between the standard model and the $R_{\rm h}=ct$ 
universe using a combination of \hii~galaxy and cosmic chronometer measurements
in the local Universe, and directly compare the results to the constraints 
imposed by the ECs. We find that the latter cosmology is not only strongly
favored by these data, with a likelihood of $\sim 92\%$ versus only $\sim 8\%$
for the former, but that its optimized fit is fully compliant with all
four ECs, while $\Lambda$CDM's best fit violates the so-called strong
energy condition at $z\lesssim 2$.}
%    \PACS{{04.20.Ex},\  {95.36.+x},\  {98.80.-k},\  {98.80.Jk}}
\maketitle

\section{Introduction}
Inflation was first proposed almost 50 years ago
\cite{1979JETPL..30..682S,1980ApJ...241L..59K,1981PhRvD..23..347G,1982PhLB..108..389L}
to resolve several conflicts between the standard model, $\Lambda$CDM,
and observations. The proposed accelerated expansion of the Universe was
meant to solve major inconsistencies, including the temperature horizon problem,
the spatial flatness in the cosmic spacetime, and the so-called monopole problem.
Inflationary cosmology gradually became the standard picture, enjoying several
notable successes, such as explaining the multi-peak structure in the angular
power spectrum of the cosmic microwave background (CMB). It also accounted rather
well for Baryonic Acoustic Oscillations and, perhaps to a lesser degree, the
polarization of the CMB.

In this picture, dark energy in the guise of a cosmological constant
accounts for the perceived current accelerated expansion of the Universe,
while cold dark matter was chiefly responsible for the formation of
large-scale structure. The scalar inflaton field, though, has never
been directly observed, and its properties are inferred only weakly
using an optimization of model parameters at much later times than
the period ($\sim$$10^{-35}$-$10^{-34}$s) when the exponentiated
expansion was supposed to have occurred.

But the latest set of more precise cosmological measurements have
begun to unravel this basic picture, which was largely constructed
empirically based on poorly sampled data several decades ago. The
newer observations from the Dark Energy Spectroscopic Instrument
\cite{2024AJ....168...58D}, the James Webb Space Telescope (JWST)
\cite{2022ApJ...936L..14P,2022ApJ...940L..55F,2022ApJ...935..110T,2024Natur.633..318C},
{\it Planck} \cite{2014A&A...571A..22P,2020A&A...641A...6P}, and
even some older Hubble Space Telescope data \cite{2016ApJ...819..129O},
are at odds with the predictions of $\Lambda$CDM.

These measurements have revealed several significant challenges to the standard
scenario, including: (i) The too early appearance of well-formed galaxies at
$z > 14$ \cite{2024PDU....4601587M}. First identified by the Hubble Space
Telescope at $z\sim 10$, this problem has been greatly exacerbated by the
James Webb Space Telescope, which discovered galaxies up to $z\sim 17$.
In the context of $\Lambda$CDM, these galaxies appear at $\sim 230$ Myr
following the Big Bang, {\it before} the emergence of
Pop~III and Pop~II stars at $\sim 280$ Myr; (ii) The discovery of
supermassive black holes at redshifts approaching $\sim 10$ \cite{Melia:2024b}.
In the context of $\Lambda$CDM, these objects also would have started forming
before the appearance of the earliest stars---and even the Big Bang. (iii) The
formation and growth of Polycyclic Aromatic Hydrocarbon grains at redshifts
$\sim 7$ that should have taken over a Gyr to form, yet were detected
at $\sim 500$ Myr in $\Lambda$CDM \cite{Melia:2024a}. Many of these conflicts
are due to the same time compression problem in the standard model, i.e.,
an age of the Universe at $z>6$ that is too short by about a factor 2.

The challenges faced by $\Lambda$CDM are not restricted to just the
observations, however. Standard cosmology is also inconsistent with
several fundamental physical principles in quantum mechanics, particle
physics, and general relativity \cite{2022PASP..134l1001M}. Each
of them is a major hurdle on its own, but when viewed as a group they
point to a need for a major overhaul of the standard theory
\cite{Melia:2020,Melia:2026}. An example
of these is  the electroweak horizon problem, in which $\Lambda$CDM
cannot account for the fact that the Higgs vacuum expectation value
is universal, beyond the causally connected regions in this standard
scenario \cite{Melia:2018a}. The electroweak symmetry would have been broken at
$t \sim 10^{-11}$ s, well beyond the hypothesized inflationary
transition at $t \sim 10^{-35}$ s, so this event creates
a second horizon problem, independent of the CMB. We now
know that the Higgs field is real and that its vacuum expectation
value is universal, so the standard model is challenged strongly
by this new evidence from particle physics. Similarly difficult
challenges emerge from the cosmic initial entropy problem and
the principle of equivalence, among several others summarized in
ref.~\cite{2022PASP..134l1001M}.

The alternative Friedmann-Lema\^itre-Robertson-Walker (FLRW) cosmology
known as the $R_{\rm h}=ct$ universe not only fits the cosmological
data better than the current standard model, but also eliminates
all of its inconsistencies with fundamental physics
\cite{Melia:2020,Melia:2026}. By now the results
of over 30 different types of comparative tests have
appeared in the literature showing that the observations favour
$R_{\rm h}=ct$ over $\Lambda$CDM, often at a high level of
confidence.

The mitigation of $\Lambda$CDM's inconsistencies includes: (i) providing
a consistent timeline for the formation of large-scale structure at
$z> 6$. In this model, the dark ages lasted until $\sim 830$ Myr after
the big bang, while the Epoch of Reionization persisted over the period
$t\subset (830-1890)$ Myr. The $5-20 M_{\odot}$ seeds that grew into high-$z$
quasars were all formed after the transition between these two periods,
consistent with the standard astrophysics of star formation in the early
Universe; (ii) an elimination of the electroweak phase transition problem.
In the $R_{\rm h}=ct$ universe, regardless of when an event (like the
electroweak phase transition) took place, the causally-connected regions
always expanded to fill the entire visible Universe today
\cite{2018EPJC...78..739M}; (iii) a removal of the tension between the
standard model's predictions and the latest measurements, based on the
AGN Hubble diagram, high-$z$ quasar Hubble diagram, the constancy of the
cluster gas mass fraction, and so on. The Bayes Information Criterion
applied to these tests typically favour the $R_{\rm h}=ct$ universe with
likelihoods exceeding $\sim 90-95\%$, compared to only $\sim 5-10\%$ for
the standard model.

At a fundamental level, the $R_{\rm h}=ct$ universe is supported
by several well-established constraints in our current physical
theories. For example, its equation of state $p=-\rho/3$
for the total pressure $p$, in terms of the total energy density
$\rho$ in the cosmic fluid, is consistent with the zero active mass
condition in general relativity, which is required for the validity
of the FLRW ansatz \cite{2022MPLA...3750016M}. As a result, this
model does not need inflation to have fixed any inconsistencies in
the early Universe. It also has no entropy problem, nor a monopole
problem. And unlike the standard model, which violates at least one
of the energy conditions from general relativity (both at high and
low redshifts, as we shall see), $R_{\rm h}=ct$ is completely
consistent with all of the classical energy conditions at all
times. And it must be said that $R_{\rm h}=ct$ universe does all
of this with just one free parameter---the Hubble Constant, $H_0$---while
$\Lambda$CDM struggles with as many as 11 or 12 parameters.

In this paper, we address one of the growing concerns with the standard
model at low redshifts, stemming from its inconsistency with the
well-established energy conditions in general relativity, which
gave rise to the FLRW metric in the first place. We carry out a joint
analysis of two independent data sets, \hii~galaxies and cosmic
chronometers, and demonstrate that not only do the latest data
strongly favour $R_{\rm h}=ct$ over $\Lambda$CDM, but they do
so satisfying the energy conditions in the former, while violating
the so-called `strong energy condition' (SEC) in the latter.

The \hii~galaxy catalog has now been extended to $z\sim 8$ by the
latest JWST discoveries \cite{2024A&A...684A..87D}. These sources
serve as standard candles because their H$\beta$ luminosity
is correlated with the ionized gas velocity dispersion where the
line radiation is produced. The relatively small scatter in these
physical quantities is due to the fact that both the number of
ionizing photons and the turbulent velocity of the gas increase
with the mass of the starburst component. In their most recent
analysis, ref.~\cite{2025MNRAS.538.1264C} used this correlation to map
the Hubble flow from redshift $\sim 0$ to $7.5$ in the context
of flat-$\Lambda$CDM, corresponding to over 12 Gyr of cosmic
expansion.

In a complementary fashion, luminous red galaxies known as
cosmic chronometers provide us with a unique way of determining
the universal expansion rate, $H(z)$, as a function of $z$.
These measurements have been compiled by directly measuring the
change in cosmic time as a function of redshift, which directly
yields the Hubble Parameter. Galaxies with an age difference
smaller than their evolutionary timescale provide the best cosmic
chronometers, whose catalog now includes over 30 measurements
\cite{2012JCAP...07..053M,2012JCAP...08..006M}.

These are the two independent data sets we shall combine for
the joint analysis. Their selection is motivated in part by
the facility with which the mathematical formalism of the
energy conditions can be implemented to gauge whether or not
the optimized fits are consistent with these general relativistic
constraints. This is an area that has received very little attention
thus far, but clearly needs to be explored more carefully going
forward. We shall analyze these data and carry out
model comparison, first by using each data set separately, and
then combining them for a joint study. As we shall see, the
optimized parameters are not completely consistent in these
three approaches, speaking to the possibility that the data 
aquisition may have missed some systematic error or simply
underestimated the overall error in certain measurements. 
We shall also find, however, that all three approaches produce
the same model selection outcome. 

The four energy conditions require that : (i) a timelike observer always
sees a non-negative energy density (the Weak Energy Condition, or
WEC); (ii) the energy density is non-negative along a null vector,
i.e., a four-vector whose time component is equal to the norm of
its spatial components (the Null Energy Condition, or NEC);
(iii) matter gravitates towards matter (the Strong Energy Condition,
or SEC); and (iv) all energy and momentum fluxes are causal and
oriented in the same way as the proper time of the observer (the
Dominant Energy Condition, or DEC).

Together with the FLRW ansatz, these four energy conditions
constrain various physical quantities, such as the pressure,
density, the expansion rate and its time derivatives. With them, 
one can develop model independent bounds on the various quantities 
(see Equations~20 to 26 below). Some previous work along these lines
has already appeared in the literature, including a demonstration
that the standard model's interpretation of Type Ia supernovae
clearly violates at least one of the energy conditions
\cite{2007PhRvD..75h3523S}.
It has also been demonstrated that inflationary cosmology
violates the energy conditions, while any basic scalar model
of expansion in the early Universe based on the zero active mass
condition does not \cite{2023AnP...53500157M}.

This paper is structured as follows: in \S2, we provide a brief
theoretical basis for the FLRW metric and the general relativistic
energy conditions applied to an FLRW universe. Following this, we
present in \S3 the mathematical constraints on the distance modulus
and Hubble parameter in an expanding Universe. In \S4, we describe
the data and methodology, and we then report the optimization
procedure and best fits for flat-$\Lambda$CDM and $R_{\rm h}=ct$ in 
\S5 and \S6.  We end with a discussion in \S7 and conclusion in \S8.

\section{Theoretical background}
\subsection{Friedmann-Lema\^itre-Robertson-Walker metric}
We shall consider the specific application of the energy conditions to the local
Universe, consistent with the Cosmological principle. The implied symmetries
reduce the general spherically-symmetric ansatz of the spacetime metric to
its Friedmann-Lema\^itre-Robertson-Walker (FLRW) form,
\begin{equation}
ds^2=c^2\,dt^2-a(t)^2\left[{dr^2\over 1-kr^2}+r^2\left(d\theta^2
+\sin^2\theta\,d\phi^2\right)\right]\,,\label{eq:FLRW}
\end{equation}
in terms of the comoving coordinates $(ct,r,\theta,\phi)$, the homogeneous expansion
factor $a(t)$, and the spatial curvature constant, $k$. The observations are consistent
with a spatially flat universe \cite{2020A&A...641A...6P} so we shall adopt the value $k=0$
throughout this paper.

Guided by the Cosmological principle, we also assume the perfect fluid approximation,
in which the stress-energy tensor, $T_{\mu\nu}$, simplifies to
\begin{equation}
T_{\mu\nu}^{0}=u_\mu u_\nu(\rho+p)/c^2-pg_{\mu\nu}\,,\label{eq:Tperfect}
\end{equation}
in terms of the 4-velocity, $u_\mu$, energy density, $\rho$, and pressure, $p$, in the fluid's
rest frame. In a perfect fluid, there are no shear forces carrying momentum components
in directions other than those associated with the components themselves \cite{Melia:2020}.
In the comoving frame, $u_\mu=(c,0,0,0)$, and so $T_{\mu\nu}={\rm diag}(\rho,p,p,p)$.
The energy density and pressure are given, respectively, as
\begin{equation}\label{PerfectFluid_Density}
\rho = \frac{3}{8\pi G}\bigg[ \frac{\dot a^2}{a^2} + \frac{k}{a^2} \bigg]\,,
\end{equation}
\begin{equation}\label{PerfectFluid_Pressure}
p = -\frac{1}{8\pi G}\bigg[ 2\frac{\ddot a^2}{a^2} + \frac{\dot a^2}{a^2} + \frac{k}{a^2} \bigg]\,,
\end{equation}
where $G$ is Newton's constant and the dots mean derivatives with respect to 
cosmic time.

Folding the FLRW metric in Equation~(\ref{eq:FLRW}) through Einstein's field equation
yields the Friedmann equation
\begin{equation}
H^2\equiv \left({\dot{a}\over a}\right)^2={8\pi G\over 3c^2}\rho-
{kc^2\over a^2}\,,\label{eq:Friedmann}
\end{equation}
the Raychaudhuri (or acceleration) equation \cite{Ray:1955},
\begin{equation}
{\ddot{a}\over a}=-{4\pi G\over 3c^2}(\rho+3p)\,,\label{eq:Ray}
\end{equation}
and the conservation of stress-energy,
\begin{equation}
\dot{\rho}=-3H(\rho+p)\,.\label{eq:encons}
\end{equation}
It is always understood that $\rho$ and $p$ in these equations include all
contributions from the cosmic fluid and the cosmological constant (if present),
so a universe dominated by matter, radiation and $\Lambda$ would be characterized
by an energy density written as $\rho=\rho_{\rm m}+\rho_{\rm r}+\rho_\Lambda$,
where $\rho_\Lambda\equiv c^4\Lambda/8\pi G$.

\subsection{The energy conditions}
The most recognizable application of the energy conditions in general
relativity, especially after the awarding of the 2020 Nobel prize in physics 
to Sir Roger Penrose \cite{Penrose:1965}, are the singularity theorems (see also
ref.~\cite{Hawking:1966,Hawking:1967,HawkingPenrose:1970,HawkingEllis:1973}).
His seminal paper showed that null geodesics inside black holes are
incomplete whenever the Ricci tensor, $R_{\alpha\beta}$, satisfies the
null convergence condition (with metric signature $[+,-,-,-]$)
\begin{equation}
R_{\mu\nu}k^\mu k^\nu\le 0\,,\label{eq:Ricci.null}
\end{equation}
for any null vector $k^\mu$. With this condition, even null geodesics
inevitably encounter a termination point, or singularity.

Equation~(\ref{eq:Ricci.null}) is a geometric constraint, but Einstein's field
equation translates it into a complementary constraint on $T^{\mu\nu}$ itself:
\begin{equation}
T_{\mu\nu}k^\mu k^\nu\ge 0\,,\label{eq:stress.null}
\end{equation}
given that $g_{\mu\nu}k^\mu k^\nu=0$.
This equation is known as the Null Energy Condition (NEC), that we now
recognize as the first of several constraints on the
energy in general relativity. Though the NEC was not used directly in the
singularity theorems, it nevertheless represents a reasonable statement
concerning the gravitational sources in $T_{\mu\nu}$ if the null convergence
condition is real \cite{Curiel:2017}.

The energy conditions we describe here are derived within {\it classical}
general relativity, and have been applied to the area increase theorem for
black hole horizons \cite{HawkingEllis:1973}, the topological censorship theorem
\cite{Friedman:1993}, and the positive mass theorem \cite{Schoen:1979}.
They are violated somewhat by quantum effects \cite{Fulling:1989}, but
these do not appear to be relevant to large classical systems, such as the
cosmological spacetime.

Note, for example, that Penrose's \cite{Penrose:1965} and Hawking and
Ellis's \cite{HawkingEllis:1973} classical singularity theorem attesting
to the existence of a Big Bang is based on the null convergence condition in
Equation~(\ref{eq:Ricci.null}), equivalent to the NEC in
Equation~(\ref{eq:stress.null}). It would be contradictory to maintain
that such a constraint was required at $t=0$, but not thereafter.
Even so, a caveat to this argument is that quantum effects could
have altered the dynamics of the Universe prior to the Planck time,
eliminating the need for a singular beginning (see, e.g.,
ref.~\cite{Visser:1995,Feldbrugge:2017}).

The pointwise energy conditions we use here restrict the form of $T^{\mu\nu}$
by ensuring that its contraction with two 4-vectors is never negative. This
ensures that $\rho$ must always be positive and that the flow of energy is never
superluminal. The four most commonly used energy conditions are the following
(see, e.g., ref.~\cite{Poisson:2007}):

\vskip 0.1in
{\bf Weak Energy Condition (WEC):} A timelike observer with 4-velocity
$u^\mu$ must always see
\begin{equation}
T_{\mu\nu}u^\mu u^\nu\ge 0\,.\label{eq:WEC}
\end{equation}
With a perfect fluid (Equation~\ref{eq:Tperfect}), this condition is equivalent to
the effective constraints
\begin{eqnarray}
\rho&\ge& 0\nonumber \\
\rho+p&\ge& 0\,.\label{eq:WECequiv}
\end{eqnarray}
So an observer with a timelike worldline always measures a non-negative
total energy density. The pressure may be positive or negative but, if
the latter, $|p|$ must never exceed $\rho$. The most concrete physical
interpretation of this energy condition (see Equation~\ref{eq:encons}) is
that the WEC prevents $\rho$ from ever increasing as the Universe expands.

\vskip 0.1in
{\bf Null Energy Condition (NEC):} This energy condition is similar to
the WEC, except that the constraint on $T_{\mu\nu}$ is imposed on a null
geodesic. The NEC condition is thus given in Equation~(\ref{eq:stress.null})
which, for a perfect fluid, is equivalent to
\begin{equation}
\rho+p\ge 0\,.\label{eq:NECequiv}
\end{equation}

\vskip 0.1in
{\bf Strong Energy Condition (SEC):} The SEC emerges from the requirement
that matter should gravitate towards matter, represented by a convergence
constraint analogous to Equation~(\ref{eq:Ricci.null}), except along the
worldline of a timelike observer \cite{Martin-Moruno:2017},
\begin{equation}
R_{\mu\nu}u^\mu u^\nu\le 0\,.\label{eq:Ricci.obs}
\end{equation}
Contracting Einstein's field equation with $g^{\mu\nu}$ yields the
Ricci scalar
\begin{equation}
R=-{8\pi G\over c^4}T\,,\label{eq:Ricci}
\end{equation}
in terms of the trace, $T\equiv T_{\mu\nu}g^{\mu\nu}$, of $T_{\mu\nu}$.
Then Einstein's field equation may be written in the form
\begin{equation}
R_{\mu\nu}=-{8\pi G\over c^4}\left(T_{\mu\nu}-{1\over 2}Tg_{\mu\nu}\right)\,,\label{eq:Einstein3}
\end{equation}
allowing us to express the strong energy condition via the equation
\begin{equation}
\left(T_{\mu\nu}-{1\over 2}Tg_{\mu\nu}\right)u^\mu u^\nu\ge 0\,.\label{eq:SEC}
\end{equation}
Its effective representation for a perfect fluid is thus
\begin{eqnarray}
\rho+p&\ge& 0\nonumber \\
\rho+3p&\ge& 0\,.\label{eq:SECequiv}
\end{eqnarray}

For the central theme in this paper, we point out that both null and timelike
geodesics must converge to a point \cite{HawkingEllis:1973} in order for a
singularity to emerge within the cosmic spacetime. Thus both
Equations~(\ref{eq:Ricci.null}) and (\ref{eq:Ricci.obs}) must be
satisfied. In other words, the existence of a Big Bang singularity
requires the Universe to respect the SEC, not just the NEC. The
latter is automatically satisfied when the SEC is also satisfied.

\vskip 0.1in
{\bf Dominant Energy Condition (DEC):}
Unlike the others, this condition encompasses two criteria: (i) the WEC and
(ii) the energy and momentum fluxes must be causal and orientated in the same
way as the proper time of the observer. The resulting two constraints are
therefore Equation~(\ref{eq:WEC}) and
\begin{equation}
{T^\mu}_\nu u^\nu {T_\mu}^\alpha u_\alpha\le 0\,,\label{eq:DEC}
\end{equation}
in which $-{T^\mu}_\nu u^\nu$ represents the 4-vector flux. With these
constraints, the energy and momentum cannot be transported superluminally.
In a perfect fluid, the DEC results in the constraints
\begin{eqnarray}
\rho&\ge& 0\nonumber \\
|p|&\le& \rho\,.\label{eq:DECequiv}
\end{eqnarray}

\subsection{Important caveats}
There is still some debate concerning the validity of the classical
energy conditions, almost always generated by the failure of favoured models
and scenarios to fully satisfy them, including inflation, dark energy in the
form of a cosmological constant, wormholes and warp drives. In truth, however,
no unassailable evidence has ever emerged that any of the models or systems
violating the energy conditions is unquestionably real or correct. Classical
energy conditions assess the level of pathology in one's choice of stress-energy
tensor, but their use is often reassessed, especially when they are violated.

Arguments questioning their validity tend to be specious and biased. At a
fundamental level, the energy conditions are certainly based on strong physical
intuition and sound theoretical justification. Take the NEC, for example. Its
violation would point to the emergence of instabilities in the system
\cite{Dubovsky:2006,Buniy:2006} and superluminal propagation
\cite{Lobo:2003}. It has been argued that the NEC must be violated
for wormholes to exist \cite{Morris:1988}, or warp drives \cite{Alcubierre:1994}.
Conventional wisdom in cosmology maintains that the SEC must be violated in
order for inflation to have occurred (see, e.g., ref.~\cite{Kontou:2020}).
The question, though, is why one should necessarily adopt the view that
the energy conditions must be flawed rather than the opposite position---that a
violation of the energy conditions argues against the validity and reality
of pathological processes.

More dramatically, a violation of the WEC would spell catastrophe
for the classical Universe because its vacuum would become unstable
to the spontaneous separation of regions with positive and negative energy.
What we do know for sure is that normal matter, e.g., baryons and
non-baryons, and relativistic particles such as photons and neutrinos,
all do satisfy every single standard energy condition.

Quantum effects do violate the classical energy conditions, but these
do not appear to be relevant on a cosmic scale. For example, the most
often used quantum argument against the classical energy conditions
is based on the Casimir effect (see, e.g., ref.~\cite{Roman:1986}),
in which the Casimir vacuum contains a negative energy density and
thus violates all of the classical energy conditions.

But consider how this negative energy state is achieved. Quantum
vacuum fluctuations of the electromagnetic field induce a small
attractive force between two close parallel uncharged conducting
plates. The classical energy conditions, however, are framed by
the dynamical evolution of an {\it unbounded} vacuum, not the
bounded system required to create the Casimir effect. The negative
energy density in the Casimir vacuum results from the boundary
conditions on the plates, quite unlike the unbounded dynamics
underlying the classical energy conditions. One cannot scale
the Casimir effect to a homogeneous and isotropic FLRW Universe
because the required boundary conditions would violate the
Cosmological principle.

Another argument raised against the validity of the energy conditions
will actually feature very prominently in this paper. The optimized
fitting of low-redshift data in the standard model, such as Type Ia supernovae,
seems to suggest that the Universe is currently accelerating. That too
violates the energy conditions. However, the existence of dark energy,
which appears to be overwhelmingly supported by the data now, does not
necessarily imply that the expansion is accelerating. It merely requires
that the expansion not be decelerating. The comparative test we shall
carry out below will directly address this very point, and our results
will demonstrate that a non-accelerating expansion actually fits the
low-redshift data better than the conventional $\Lambda$CDM. Thus,
one cannot insist that the acceleration implied by the standard model
argues against the validity of the energy conditions.

\section{Constraints from the energy conditions}
Writing the energy constraints in terms of the time derivatives of the expansion
rate of the Universe using Equations~\eqref{PerfectFluid_Density} and
\eqref{PerfectFluid_Pressure}, we get:
\begin{equation}\label{NEC}
\hskip-1.0in\textbf{NEC:}\quad \:\: -\frac{\ddot a}{a} + \frac{\dot a^2}{a^2} \geq 0\,,
\end{equation}
\begin{equation}\label{SEC}
\hskip-1.55in\textbf{SEC:}\;\;\;\qquad \:\: \frac{\ddot a}{a} \leq 0\,,
\end{equation}
\begin{equation}\label{WEC}
\hskip-1.4in\textbf{WEC:}\qquad \:\: \frac{\dot a^2}{a^2} \geq 0\,,
\end{equation}
and
\begin{equation}\label{DEC}
\hskip-0.7in\textbf{DEC:}\qquad \:\: -\frac{\dot a^2}{a^2} \leq -\frac{\ddot a}{a} \leq
2\frac{\dot a^2}{a^2}\,.
\end{equation}
On integrating these once, we find that:
\begin{equation}\label{NEC_Use}
\hskip-1.32in\textbf{NEC:}\qquad \:\: \dot a \geq a H_{0}\,,
\end{equation}
\begin{equation}\label{SEC_Use}
\hskip-1.28in\textbf{SEC:}\;\qquad \:\: \dot a \geq a_{0} H_{0}\,,
\end{equation}
and
\begin{equation}\label{DEC_Use}
\hskip-1.22in\textbf{DEC:}\qquad \:\: \dot a \leq \frac{a_0^3}{a^2} \:  H_{0}.
\end{equation}

\subsection{EC bounds on the distance modulus}
The luminosity distance $D_L(z)$, in terms of the acceleration of the
Universe, is given by
\begin{equation}\label{D_L}
D_L(z) = a_{0}(1+z)\:d(a),
\end{equation}
where the comoving distance $d(a)$, for a flat universe, is given by
\begin{equation}\label{RadialDistance_Definition}
d(a) = \int_{a}^{a_0=1}{\frac{1}{\dot aa} da} .
\end{equation}
Hence, the bounds on the distance modulus for different energy conditions using
Equations~\eqref{NEC_Use}--\eqref{DEC_Use}, \eqref{D_L},
\eqref{RadialDistance_Definition} and (\ref{distmod}) are
\begin{equation}\label{DistanceModulus_NEC_UpperBound}
\textbf{NEC:} \:\: \mu(z) \leq 5 \log_{10} \bigg[ \frac{z(1+z)}{H_0} \bigg] + 25,
\end{equation}
\begin{equation}\label{DistanceModulus_SEC_UpperBound}
\textbf{SEC:} \:\: \mu(z) \leq 5 \log_{10} \bigg[ \frac{(1+z)}{H_0} \ln(1+z) \bigg] + 25,
\end{equation}
\begin{equation}\label{DistanceModulus_DEC_LowerBound}
\textbf{DEC:} \:\: \mu(z) \geq 5 \log_{10} \bigg[ \frac{(1+z)}{2H_0}
\bigg(1-\frac{1}{(1+z)^2} \bigg) \bigg] + 25.
\end{equation}

\subsection{EC bounds on the Hubble parameter}
The bounds on the Hubble Parameter for different energy conditions
using Equations~\eqref{NEC_Use}--\eqref{DEC_Use} and (\ref{HubbleParameter_Definition}) are
\begin{equation}\label{HubbleParameter_NEC_LowerBound}
\textbf{NEC:} \:\: H \geq H_0,
\end{equation}
\begin{equation}\label{HubbleParameter_SEC_LowerBound}
\textbf{SEC:} \:\: H \geq H_0(1+z),
\end{equation}
\begin{equation}\label{HubbleParameter_DEC_UpperBound}
\textbf{DEC:} \:\: H \leq H_0(1+z)^3.
\end{equation}

\begin{table*}
\begin{center}
\caption{Flux and gas velocity dispersion of \hii~galaxies discovered by JWST$^{a}$.}\label{table1}
\begin{tabular}{lccc}
&&& \\
\hline\hline
Name&\emph{z}&$\log \sigma_v$&$\log F(\hb)$\\
    &        &$(\mathrm{km\,s^{-1}})$&$(\mathrm{erg\,s^{-1}\,cm^{-2}})$\\
\hline
%&&& \\
JADES-NS--00016745 & 5.56616  $\pm$ 0.00011  & 1.731 $\pm$  0.016 &  $-17.64 \pm  0.21$ \\
JADES-NS--10016374 & 5.50411  $\pm$ 0.00007  & 1.785 $\pm$  0.014 &  $-18.14 \pm  0.20$ \\
JADES-NS--00019606 & 5.88979  $\pm$ 0.00008  & 1.622 $\pm$  0.019 &  $-18.11 \pm  0.27$ \\
JADES-NS--00022251 & 5.79912  $\pm$ 0.00007  & 1.621 $\pm$  0.011 &  $-17.86 \pm  0.11$ \\
JADES-NS--00047100 & 7.43173  $\pm$ 0.00015  & 1.868 $\pm$  0.024 &  $-17.62 \pm  0.39$ \\
%&&& \\
\hline\hline
\end{tabular}
\begin{description}
\item[$^{a}$] {Taken from ref.~\cite{2025MNRAS.538.1264C}.}
\end{description}
\end{center}
\end{table*}

\section{Observational data and methodology}
\subsection{\hii~galaxies}
\medskip
The data are taken from ref.~\cite{2025MNRAS.538.1264C}, including
a full sample of 181 HIIGx analyzed in ref.~\cite{2021MNRAS.505.1441G},
9 newer HIIGx from ref.~\cite{2023A&A...676A..53L}, and the 5 HIIGx newly
discovered by JWST ref.~\cite{2024A&A...684A..87D}, constituting
a total sample of 195 independently measured sources.

We calculate the $\hb$ luminosity from the reddening corrected $\hb$ fluxes,
$F(\mathrm{H}\beta)$, and their $1\sigma$ uncertainties, as described in
ref.~\cite{2015MNRAS.451.3001T} (see also ref.~\cite{2000ApJ...533..682C}):
\begin{equation}\label{eq:L}
 L(\hb) = 4 \pi  D_L^2(z) F(\hb)\,.
\end{equation}
The most interesting of these sources are the high-redshift objects
discovered recently by JWST. We list their corrected dispersions
and $1\sigma$ uncertainties in column (3) of Table~\ref{table1}, along with
the measured $\hb$ fluxes and their errors in column (4). The rest of the catalog
may be found in ref.~\cite{2025MNRAS.538.1264C} and references cited therein.

The emission-line luminosity is correlated with the ionized gas velocity dispersion,
$\sigma_v$, \cite{2012MNRAS.425L..56C,2014MNRAS.442.3565C,2015MNRAS.451.3001T} according to
the ansatz
\begin{equation}\label{eq:L-sigma}
\log L(\hb)=\beta \log \sigma_v+\alpha\;,
\end{equation}
where $\beta$ is the slope and $\alpha$ is a constant representing the logarithmic
luminosity at $\log \sigma_v=0$. Note, though, that $\alpha$ and $\beta$ are
cosmology-dependent nuisance parameters, so they must be optimized 
simultaneously with the rest of the cosmological parameters (if any). Most cosmological
measurements, most famously of Type Ia SNe, must be made in the context of a
pre-selected model. The implied reliance on such so-called nuisance parameters is
thus unavoidable. One must therefore carefully take into account that any optimized
model parameters are therefore also dependent on these correlation variables. A
particularly relevant example, as we shall see, is the Hubble constant, which will
have different values in the models being compared. In other words, not all
cosmological parameters can be measured in a model-independent way.

With Equations~(\ref{eq:L}) and (\ref{eq:L-sigma}), the distance modulus may be written as
\begin{equation}
\mu_{\rm obs}=2.5\left[\beta \log \sigma_v+\alpha - \log F(\hb)\right]-100.2 \;,
\end{equation}
with a corresponding error ($\sigma_{\mu_{\rm obs}}$) found via error propagation,
\begin{equation}\label{eq:sigma}
\sigma_{\mu_{\rm obs}}=2.5\left(\beta^{2} \sigma^{2}_{\log \sigma_v}+\sigma^{2}_{\log F}\right)^{1/2} \;,
\end{equation}
in terms of the $1\sigma$ uncertainties, $\sigma_{\log \sigma_v}$ and $\sigma_{\log F}$,
in $\log \sigma_v$ and $\log F(\hb)$, respectively.

This measured quantity will be compared with the theoretically predicted distance modulus,
\begin{equation}
\mu_{\rm th}\equiv5 \log\left[\frac{D_{L}(z)}{\rm Mpc}\right]+25\,.\label{distmod}
\end{equation}
In $\Lambda$CDM, $D_L(z)$ is given as
\begin{eqnarray}\label{eq:DL_LCDM}
&\null& D_{L}^{\Lambda {\rm CDM}}(z) = {c\over
H_{0}}\,{(1+z)\over\sqrt{\mid\Omega_{\rm k}\mid}}\; {\rm
sinn}\Biggl\{\mid\Omega_{\rm k}\mid^{1/2}\times\nonumber\\
&\null&\hskip-0.1in \int_{0}^{z}{dz\over\sqrt{\Omega_{\rm m}(1+z)^{3}+\Omega_{\rm k}(1+z)^{2}+
\Omega_{\rm de}(1+z)^{3(1+w_{\rm de})}}}\Biggr\},\qquad
\end{eqnarray}
in terms of the energy density, $\rho=\allowbreak\rho_r+\nobreak\rho_m+\nobreak\rho_{\rm de}$,
including radiation, matter (luminous and dark), and dark energy, expressed as fractions
of today's critical density, $\rho_c\equiv 3c^2 H_0^2/8\pi G$: $\Omega_m\equiv\rho_m/\rho_c$,
$\Omega_r\equiv\rho_r/\rho_c$, and $\Omega_{\rm de} \equiv \rho_{\rm de}/\rho_c$. Also,
$p_{\rm de}=w_{\rm de}\rho_{\rm de}$ is the dark-energy equation of state and
Equation~(\ref{eq:DL_LCDM}) assumes negligible radiation pressure up to $z\sim 8$.
In this paper, we assume spatial flatness throughout our analysis, so $\Omega_{k}=0$
and the right side of this equation then reduces to $(1+z)c/H_{0}$ times the integral.

In the $R_{\rm h}=ct$ universe
\cite{2003eisb.book.....M,2007MNRAS.382.1917M,2009IJMPD..18.1889M,2012MNRAS.419.2579M},
we have instead
\begin{equation}\label{DL_Rh}
D_L^{R_{\rm h}=ct}(z)={c\over H_0}(1+z)\ln(1+z)\;.
\end{equation}

\subsection{Cosmic chronometers}
The cosmic chronometers are a set of 34 mutually independent observations of galaxies 
which provide a measurement of the differential age of the Universe, avoiding 
possible problems with integrated histories over a period where
the sources may be evolving. The expansion rate is inferred from
\begin{equation}\label{HubbleParameter_Definition}
H(z) = \frac{\dot a}{a} = - \frac{1}{1+z} \frac{dz}{dt}\,.
\end{equation}

For various reasons, the best cosmic chronometers are galaxies that evolve passively
on a timescale much longer than their age difference. Observations indicate that
the most massive galaxies contain the oldest stellar populations up to redshifts
$z \sim 1-2$
\cite{1996Natur.381..581D,1997ApJ...484..581S,1999AJ....118..603C,2004Natur.428..625H,2005ApJ...621..673T}.
More than 99\% percent of the stellar mass in these massive galaxies formed at
$z \geq 1$ \cite{2004Natur.428..625H,2007MNRAS.378.1550P}. In high-density
regions (such as galaxy clusters), star formation terminated by redshift $z \sim 3$
\cite{2005ApJ...621..673T}, and other massive systems---those with stellar
masses $5 \times 10^{11} M_{\odot}$---concluded their star formation campaigns
by $z \sim 2$ \cite{2005ApJ...633..174T}.

The empirical evidence suggests that galaxies in the highest density regions of
clusters have been evolving passively since forming their stellar content at
$z \geq 2$, without any subsequent chapters of star formation. One can therefore
think of these galaxies as tracing the `red envelope', hosting the 
oldest stars in the Universe at every redshift.

\begin{figure}[t]
\centerline{\includegraphics[angle=0,scale=0.65]{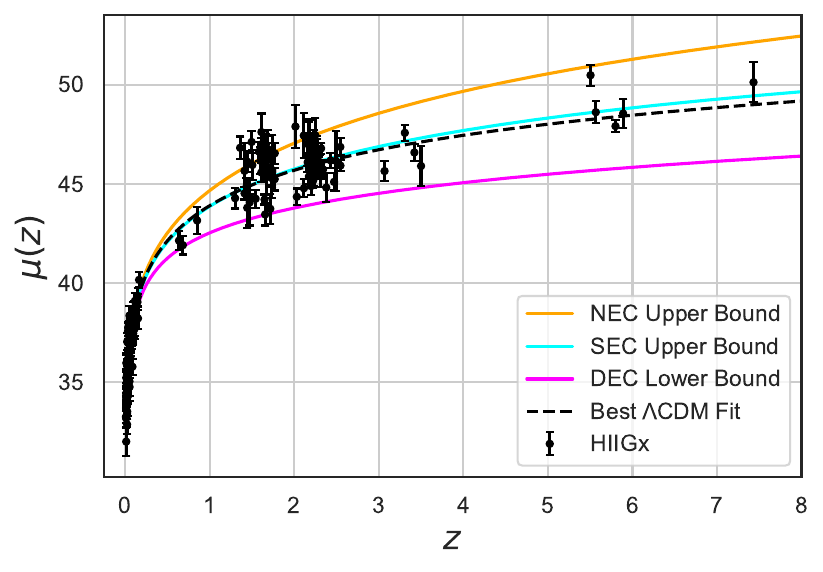}}
\caption{Distance modulus for the \hii~galaxy data in the optimized, flat
$\Lambda$CDM model. The violation of the SEC limit (blue curve) by the standard 
model prediction (dashed curve) becomes apparent in the redshift range $z\subset 
(0.2, 1.5)$, as highlighted by the magnified view in Figure~\ref{fig.2}.}\label{fig.1}
\end{figure}

\begin{figure}
\centerline{\includegraphics[angle=0,scale=0.4]{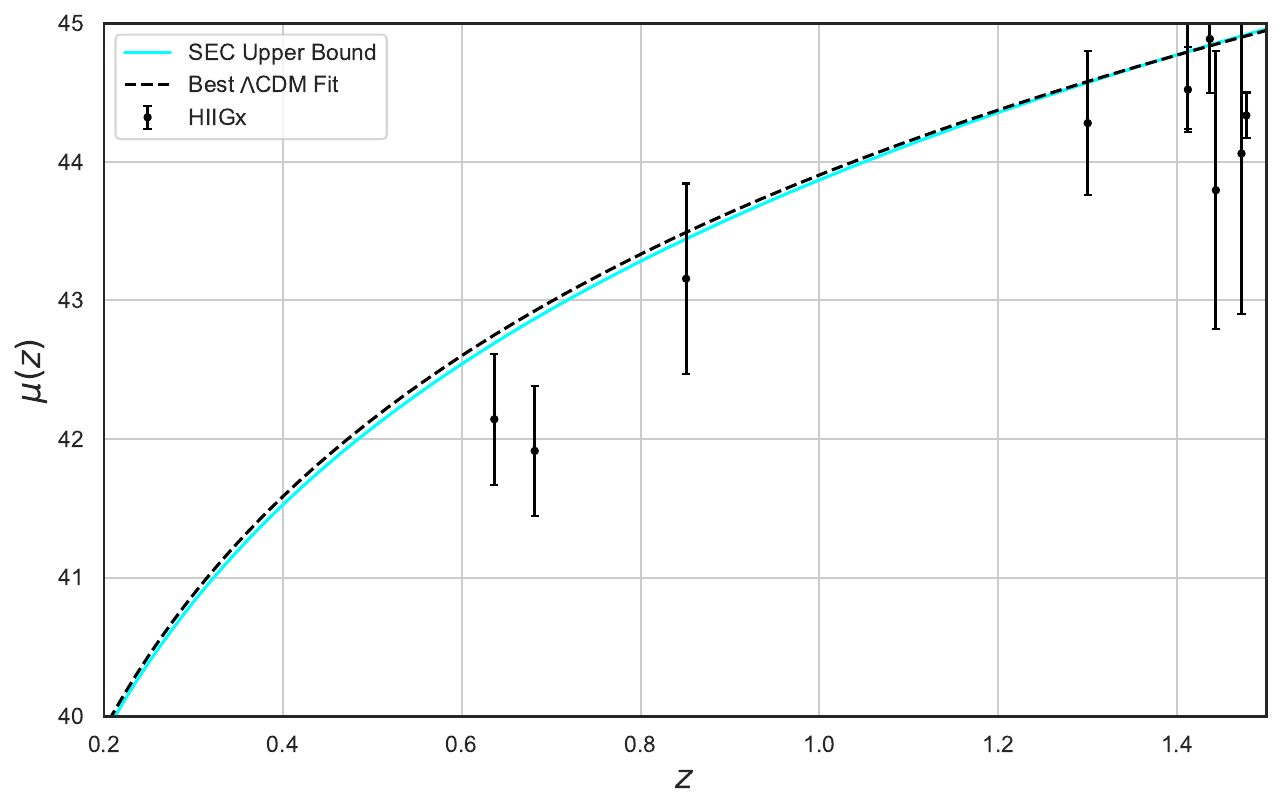}}
\caption{The magnified region $z\subset(0.2,1.5)$ from Figure~\ref{fig.1} where
the SEC violation becomes apparent.}\label{fig.2}
\end{figure}

A direct way of determining the age of these galaxies is to use the $4000 Å$ break
in its spectrum, which relates linearly to the age of old stellar populations
\cite{2011JCAP...03..045M}. This discontinuity of the spectral continuum
is caused by metal absorption lines whose amplitude correlates linearly with the age
and metal abundance. If the metallicity is known, then the difference in age between
two galaxies is proportional to the difference in their $4000 Å$ amplitudes.

One must also be aware of the many systematic errors, however, that can bias this kind
of analysis (see, e.g., ref.~\cite{2012JCAP...07..053M}) like: (i) the degeneracy
between the effect caused by a change in age and an effect due to a change in its stellar
metallicity; (ii) the choice of model for stellar population synthesis, being used to
estimate the age or calibrate the $4000 Å$ break versus age relation leading to biasing
in $H(z)$; and (iii) the possible existence of a progenitor bias
\cite{1996MNRAS.281..985V}, where the high-$z$ samples of early-type
galaxies might not be statistically equivalent to those at low redshifts.

Despite these caveats, one is motivated by the agreement between the results using different
techniques. These data were assembled from the compilations of
ref.~\cite{2005PhRvD..71l3001S,2010ApJS..188..280S,2012JCAP...08..006M},
over the redshift range $0 \leq z \leq  1.8$, providing a reasonably consistent picture of the
universal expansion, especially when viewed with reference to theoretical expectations,
that we consider in this work.

\begin{figure}
\centerline{\includegraphics[angle=0,scale=0.48]{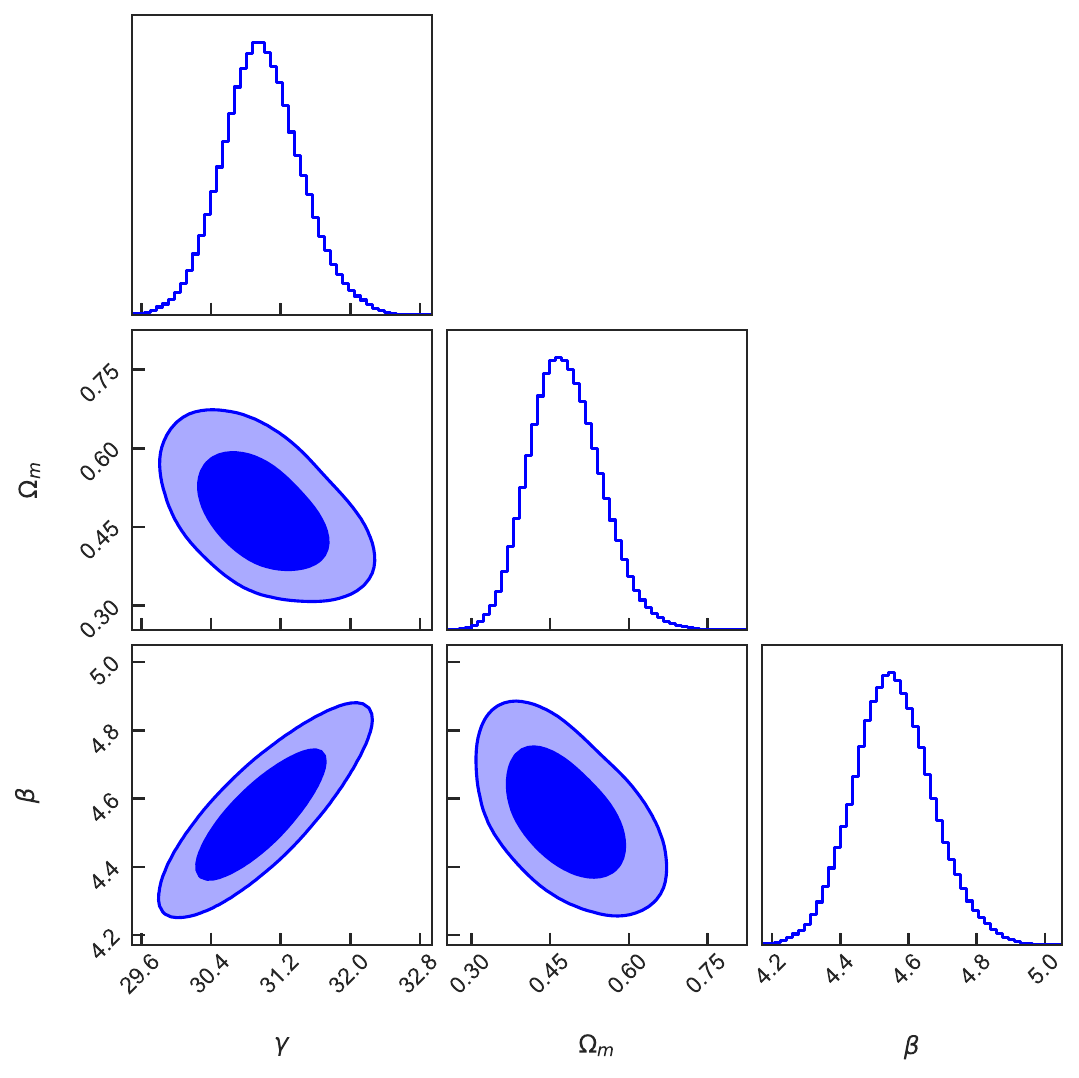}}
\caption{1D probability distributions and 2D regions with the $1-2\sigma$
contours corresponding to the parameters $\gamma$, $\Omega_{\rm m}$, and $\beta$
in the flat $\Lambda$CDM fit (fig.~\ref{fig.1}) to the \hii~galaxy data.}\label{fig.3}
\end{figure}

\section{Best fit models}
We carry out model selection comparing two distinct and unnested 
cosmologies, flat-$\Lambda$CDM and
$R_{\rm h}=ct$, using a Maximizing Likelihood Estimate Procedure, utilizing a uniform
prior for every parameter. The model parameters are optimized for each model and
dataset. The optimizations are done using the Python Markov chain Monte Carlo (MCMC)
module, EMCEE \cite{2013PASP..125..306F}, and we discuss the results for each dataset
and cosmological model in the following subsections.

\begin{figure}
\centerline{\includegraphics[angle=0,scale=0.65]{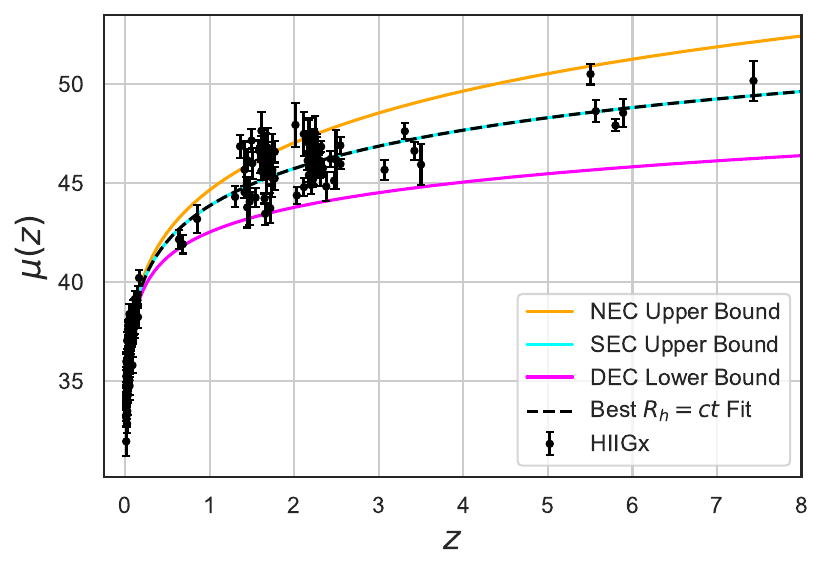}}
\caption{Same as Figure~\ref{fig.1}, except now for the $R_{\rm h}=ct$ universe. In
this case, the SEC (blue) curve is identical to the model's prediction, so this
cosmology is fully consistent with all of the energy conditions.}\label{fig.4}
\end{figure}

\begin{figure}
\centerline{\includegraphics[angle=0,scale=0.69]{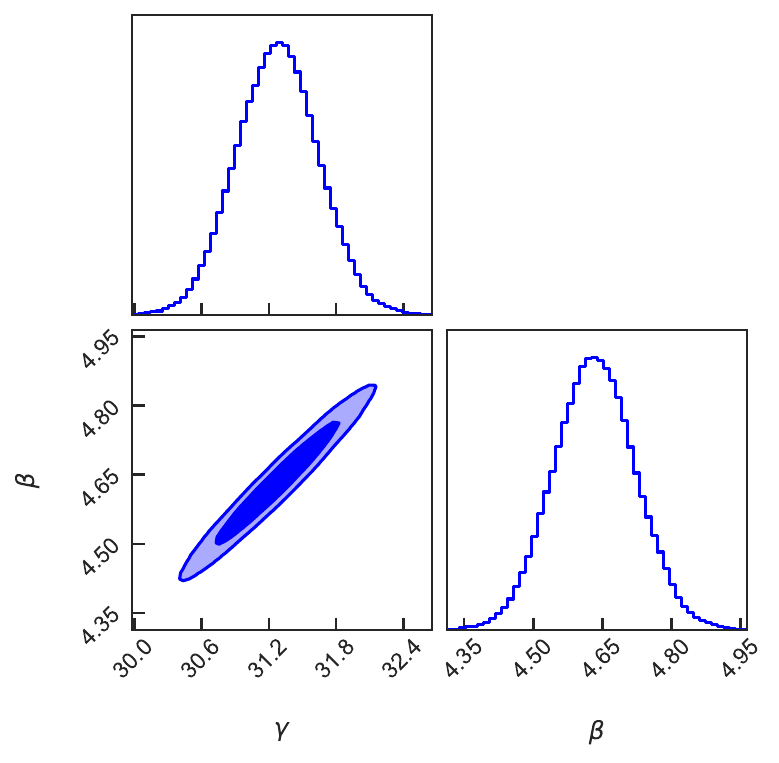}}
\caption{1D probability distributions and 2D region with the $1-2\sigma$
contours corresponding to the parameters $\gamma$, and $\beta$ in the
$R_{\rm h}=ct$ universe fit (fig.~\ref{fig.4}) to the \hii~galaxy data.}\label{fig.5}
\end{figure}

\subsection{Optimization using only \hii~galaxies}
The coefficients $\alpha$ and $\beta$, and the cosmological parameters ($H_0$, and
$\Omega_{\rm m}$ for $\Lambda$CDM and just $H_0$ for $R_{\rm h}=ct$) are optimized simultaneously
using the 195 HIIGx objects, based on an implementation of Maximum Likelihood Estimation.
The likelihood function is given by,
\begin{equation}\label{eq:likelihood1}
\mathcal{L}_{\rm HIIGx} = \prod_{i}^{195}
\frac{1}{\sqrt{2\pi}\,\epsilon_{{\rm HIIGx},i}}\;\times
\exp\left[-\,\frac{\left(\mu_{{\rm obs},i}-\mu_{\rm
th}(z_i)\right)^{2}}{2\epsilon^{2}_{{\rm HIIGx},i}}\right]\,,
\end{equation}

\begin{table*}
\begin{center}
{\footnotesize
\caption{Best-fitting results using solely the \hii~galaxy data.}\label{table2}
\begin{tabular}{lcccccccc}
&&&&&&& \\
\hline\hline
%&&&&&&& \\
Model& $\gamma$ & $\Omega_{\rm m}$ & $\beta$ & $-2\ln \mathcal{L}$ &\quad BIC & Probability\\
&&&$[\mathrm{km\,s^{-1}\,Mpc^{-1}}]$ &&&& \\
\hline
%&&&&&&& \\
$R_{\rm h}=ct$ & $31.27\pm 0.36$ & --  & $4.63\pm0.09$ & 795.46 & 806.01 & $88.36\%$ \\
%&&&&&&& \\
$\Lambda$CDM & $30.94\pm0.44$ & $0.48\pm0.07$ & $4.55\pm 0.11$ & 794.24 & 810.06 &  $11.64\%$ \\
%&&&&&&& \\
\hline\hline
\end{tabular}
}
\end{center}
\end{table*}

\noindent where the variance on each HIIGx is given by,
\begin{equation}\label{eq:variance1}
\epsilon^{2}_{{\rm HIIGx},i}=\sigma^{2}_{\mu_{\rm obs},i}+
\left[\frac{5\sigma_{D^{\rm th}_{L},i}}{\ln{10}\,D^{\rm th}_{L}(z_i)}\right]^2\,,
\end{equation}
in terms of $\sigma_{D_L^{th}, i}$---the propagated uncertainty of $D^{\rm th}_{L}(z_i)$.
Note that the first factor, $\epsilon_{\rm HIIGx}$, in Equation~(\ref{eq:likelihood1})
is not a constant, since it depends on the value of $\beta$. Thus, maximizing
${\mathcal{L}}$ is not exactly equivalent to minimizing the $\chi^{2}$ statistic,
i.e., $\chi^{2}=\sum_i\frac{\left(\mu_{{\rm obs},i} - \mu_{\rm th}(z_i)\right)^2}
{\epsilon^{2}_{{\rm HIIGx},i}}$. In addition, $-2\ln \mathcal{L} =
\chi^2 + \ln (2\pi\,\epsilon_{\rm HIIGx})$.

When using solely the \hii~galaxy data, the parameters $H_0$ and $\alpha$ are not
independent, so we re-parameterize $H_0$ and $\alpha$ as follows:
\begin{equation} \label{H0_alpha_reparametrization}
\gamma = 125.2 - 2.5\alpha - 5\log H_0\,.
\end{equation}
This re-parameterization becomes redundant when using the additional 36 \hii~sources in the
`anchor samples' (as used in ref.~\cite{2025MNRAS.542L..19W}), since the luminosities
assigned to this set constrain $\alpha$ and $\beta$ individually. We are not including
this sub-sample here because it is biased by the introduction of additional data not
available to the rest of the \hii~galaxies in our sample.
With the re-parameterization in \eqref{H0_alpha_reparametrization},
the likelihood function \eqref{eq:likelihood1} takes the following form:
\begin{equation}\label{eq:likelihood1_updated}
\mathcal{L}_{\rm HIIGx} = \prod_{i}^{195}
\frac{1}{\sqrt{2\pi}\,\epsilon_{{\rm HIIGx},i}}\;\times
\exp\left[-\,\frac{\Gamma_i^{2}}{2\epsilon^{2}_{{\rm HIIGx},i}}\right]\,,
\end{equation}
where,
\begin{equation}
\Gamma_i = 2.5\left[\beta \log \sigma_{v,i} - \log F( H\beta)_i\right] -
\gamma - 5\log(H_0 D_L^{th}[z_i]),
\end{equation}
and the variance is still given by \eqref{eq:variance1} since it does not depend on the
Hubble constant $H_0$ or $\alpha$.

\subsubsection{$\Lambda$CDM}
\medskip
In the most basic $\Lambda$CDM model, $w_{\rm de}=-1$, leaving
the two coefficients $\gamma$ (re-parameterized from $\alpha$ and $H_0$ in
Eq.~\ref{H0_alpha_reparametrization}) and $\beta$, and the matter density
$\Omega_{\rm m}$ as the free parameters. The optimized fit is shown in
Figures~\ref{fig.1} and \ref{fig.2}, and the corresponding parameter values
are presented in Table~\ref{table2}, with the $1-2\sigma$ confidence regions
plotted in Figure~\ref{fig.3}.

\begin{figure}
\centerline{\includegraphics[angle=0,scale=0.62]{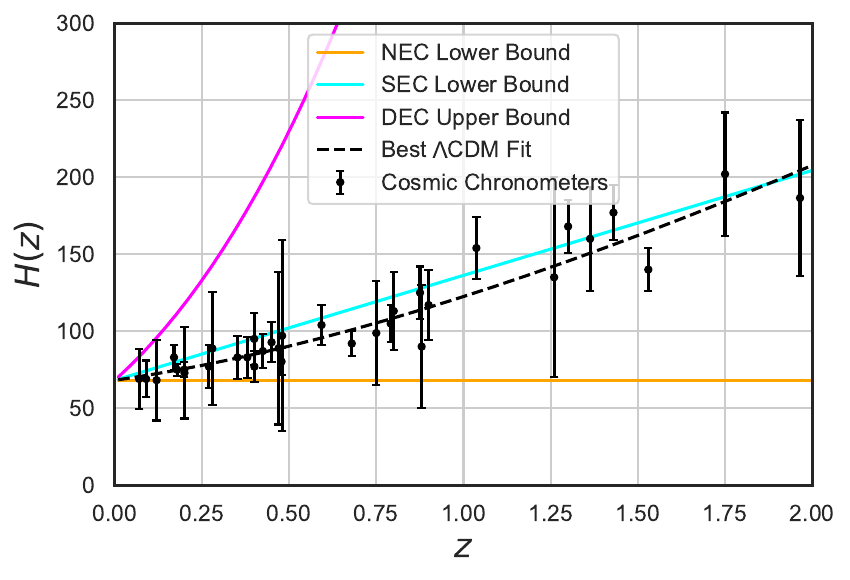}}
\caption{The optimized flat-$\Lambda$CDM fit to the cosmic chronometer data.
In this case, the violation of the SEC by the standard model can 
be seen within the redshift range $z\subset (0,1.75)$, confirming the result 
for the \hii~data in Figures~\ref{fig.1} and \ref{fig.2}.}\label{fig.6}
\end{figure}

\begin{figure}
\centerline{\includegraphics[angle=0,scale=0.68]{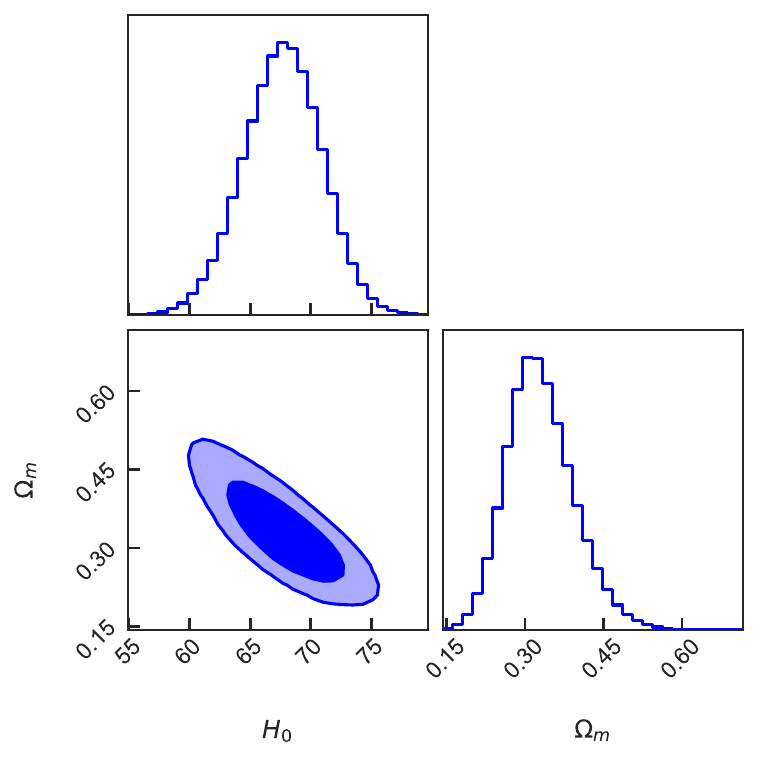}}
\caption{1D probability distributions and 2D region with the $1-2\sigma$
contours corresponding to the parameters $H_0$ and $\Omega_m$ in the
flat-$\Lambda$CDM fit (fig.~\ref{fig.6}) to the cosmic chronometer data.}\label{fig.7}
\end{figure}

\begin{figure}
\centerline{\includegraphics[angle=0,scale=0.63]{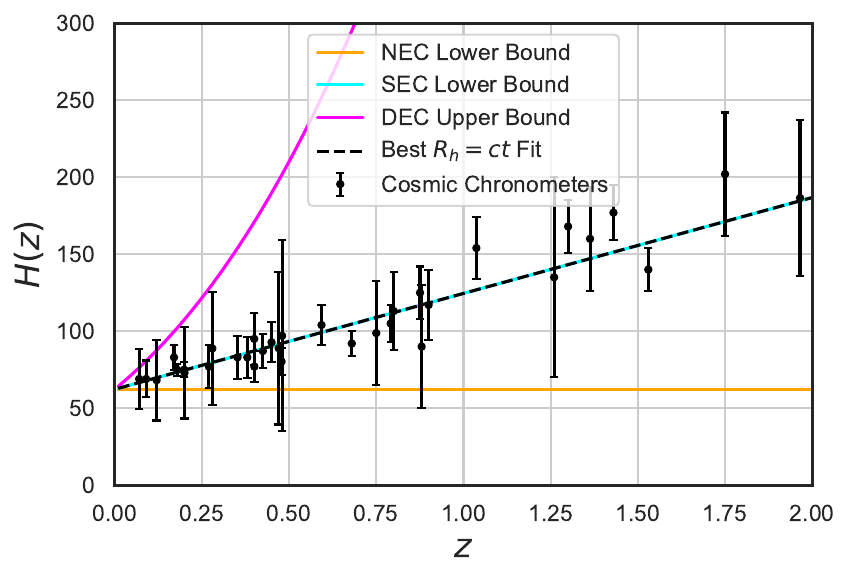}}
\caption{Same as Figure~\ref{fig.6}, except now for the $R_{\rm h}=ct$ universe.
The SEC curve is identical to the model prediction in this cosmology,
so the optimized fit to the cosmic chronometer data in $R_{\rm h}=ct$
satisfies all of the energy conditions.}\label{fig.8}
\end{figure}

\begin{figure}
\centerline{\hskip0.3in\includegraphics[angle=0,scale=1]{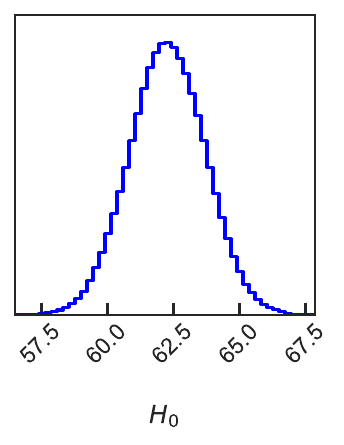}}
\vskip-0.05in
\caption{1D probability distribution corresponding to the only parameter, $H_0$, in
the $R_{\rm h}=ct$ universe fit (fig.~\ref{fig.8}) to the cosmic chronometer data.}\label{fig.9}
\end{figure}

\subsubsection{The $R_{\rm h}=ct$ universe}
\medskip
By comparison, the $R_{\rm h}=ct$ universe has only one free parameter, $H_{0}$, apart
from the two coefficients $\alpha$, and $\beta$, so after re-parameterization,
the parameters to be optimized are $\gamma$, and $\beta$. The optimized fit is shown
in Figure~\ref{fig.4}, with the parameters listed in Table~\ref{table2} and
the $1-2\sigma$ confidence regions shown in Figure~\ref{fig.5}.

\begin{figure}
\centerline{\includegraphics[angle=0,scale=0.64]{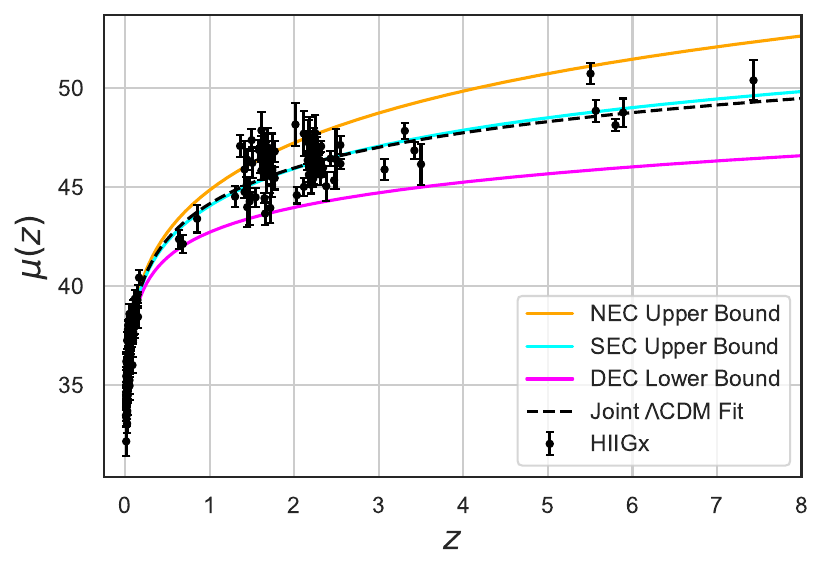}}
\caption{The flat-$\Lambda$CDM fit to the \hii~galaxy data based on the
maximum likelihood estimation of the parameters using the combined
\hii~galaxy and cosmic chronometer measurements. The standard model's violation
of the SEC, seen at intermediate redshifts in Figures~\ref{fig.1},
\ref{fig.2} and \ref{fig.6}, persists with the joint analysis.}\label{fig.10}
\end{figure}

\begin{figure}
\centerline{\includegraphics[angle=0,scale=0.62]{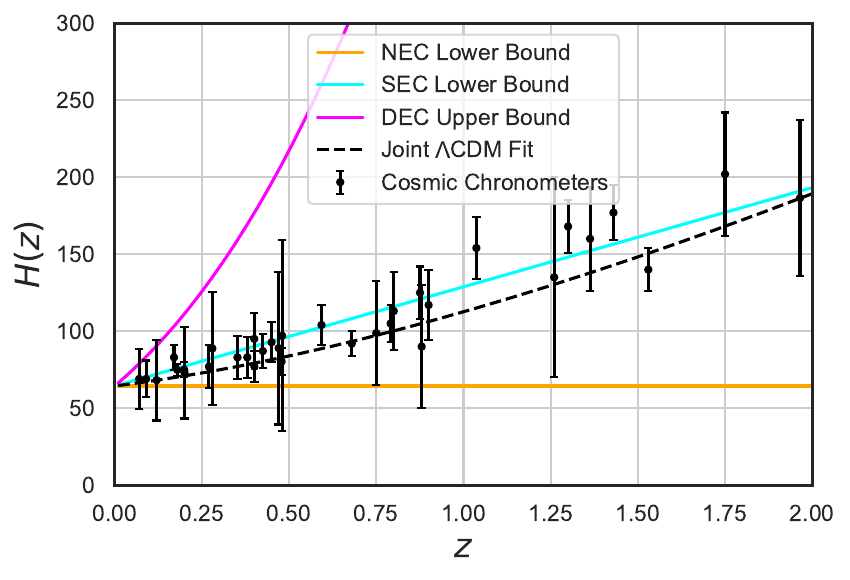}}
\caption{The flat-$\Lambda$CDM fit to the cosmic chronometer
data based on the maximum likelihood estimation of the parameters
using the combined \hii~galaxy and cosmic chronometer data. Based
on this joint optimization, the SEC violation by the standard model can again be seen
at $z\subset (0,2)$.}\label{fig.11}
\end{figure}

\begin{figure}
\centerline{\includegraphics[angle=0,scale=0.37]{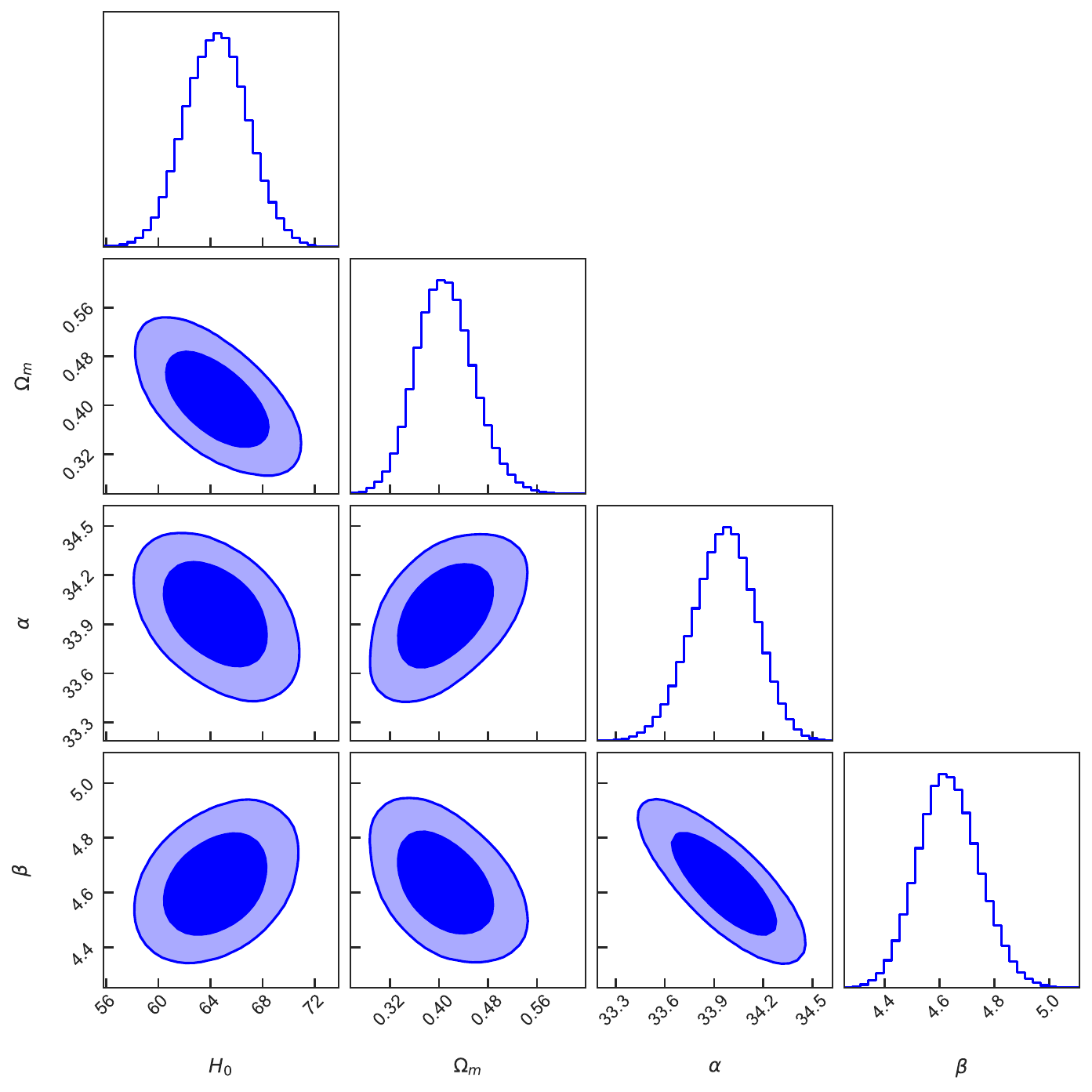}}
\caption{1D probability distributions and 2D regions with the $1-2\sigma$
contours corresponding to the parameters $\alpha$, $\beta$, $H_0$, and
$\Omega_m$ in the flat-$\Lambda$CDM joint fit (figs.~\ref{fig.10} and \ref{fig.11})
to the \hii~galaxy $+$ cosmic chronometer data.}\label{fig.12}
\end{figure}

\begin{table*}
\begin{center}
{\footnotesize
\caption{Best-fitting results using solely the cosmic chronometers.}\label{table3}
\begin{tabular}{lccccc}
&&&&& \\
\hline\hline
%&&&&& \\
Model& $H_0$ & $\Omega_{\rm m}$ & $2\ln \mathcal{L}$ &\quad BIC & Probability\\
&$[\mathrm{km\,s^{-1}\,Mpc^{-1}}]$ &&&& \\
\hline
%&&&&& \\
$R_{\rm h}=ct$            & $62.26\pm 1.42$ & --  & 277.86 & 281.38 & $65.87\%$ \\
%&&&&& \\
$\Lambda$CDM              & $67.72\pm 3.06$ & $0.33\pm 0.06$ & 275.65 & 282.70 &  $34.13\%$ \\
%&&&&& \\
\hline\hline
\end{tabular}
}
\end{center}
\end{table*}

\subsection{Optimization using only cosmic chronometers}
%\medskip
Here, the cosmological parameters are optimized using the sample of 34 cosmic-chronometer
measurements, with a Maximum likelihood Estimation given by
\begin{equation}\label{eq:likelihood2}
\mathcal{L}_{\rm cc} = \prod_{i}^{34}
\frac{1}{\sqrt{2\pi}\,\epsilon_{{\rm cc},i}}\;\times
\exp\left[-\,\frac{\left(H_{{\rm obs},i}-H_{\rm
th}(z_i)\right)^{2}}{2\epsilon^{2}_{{\rm cc},i}}\right]\,,
\end{equation}
where the variance on each cosmic chronometer is given by
\begin{equation}\label{eq:variance2}
\epsilon^{2}_{{\rm cc},i}=\sigma^{2}_{H_{\rm obs},i},
\end{equation}

\subsubsection{$\Lambda$CDM}
\medskip
For flat-$\Lambda$CDM, the 2 parameters optimized are
$H_0$ and $\Omega_m$, producing the fit shown in Figure~\ref{fig.6}.
The parameter values are listed in Table~\ref{table3}, with the $1-2\sigma$
confidence regions shown in Figure~\ref{fig.7}.

\begin{table*}
\begin{center}
{\footnotesize
\caption{Best-fitting results for the combined analysis of the \hii~galaxy
and cosmic chronometer data.}\label{table4}
\begin{tabular}{lccccccc}
&&&&&& \\
\hline\hline
%&&&&&& \\
Model& $H_0$ & $\Omega_{\rm m}$ & $\alpha$ & $\beta$ & $2\ln \mathcal{L}$ &\quad BIC & Probability\\
&$[\mathrm{km\,s^{-1}\,Mpc^{-1}}]$ & \\
\hline
%&&&&&& \\
$R_{\rm h}=ct$ & $62.17\pm1.42$ & -- & $33.98\pm 0.14$ & $4.63\pm0.09$  & $1073.32$
& $1089.62$  & $92.32\%$ \\
%&&&&&& \\
$\Lambda$CDM   & $64.26\pm2.28$ & $0.41\pm0.05$ & $33.96\pm0.19$ & $4.64\pm0.11$ & $1072.86$
& $1094.59$ &  $7.68\%$ \\
%&&&&&& \\
\hline\hline
\end{tabular}
}
\end{center}
\end{table*}

\subsubsection{The $R_{\rm h}=ct$ universe}
The $R_{\rm h}=ct$ universe has only one free parameter, $H_{0}$. The best fit
curve is shown in Figure~\ref{fig.8}, with the optimized parameter value listed
in Table~\ref{table3}, and shown in Figure~\ref{fig.9}.

\section{Optimization using the combined data sets} \label{Section 6}
The cosmological parameters and the coefficients $\alpha$ and $\beta$ are
optimized simultaneously using the 195 HIIGx and 34 cosmic chronometers based on
the procedure outlined in \S5, but with one change: we maximize the sum of the
individual likelihoods using
\begin{equation}\label{TotalLogLikelihood}
\ln(\mathcal{L}_{\rm tot})=\ln(\mathcal{L}_{\rm HIIGx}) + \ln({\mathcal{L}_{\rm cc}})\,,
\end{equation}
where $\mathcal{L}_{\rm HIIGx}$ and $\mathcal{L}_{\rm cc}$ are given by
Equations~(\ref{eq:likelihood1}) and (\ref{eq:likelihood2}), respectively.

\subsection{$\Lambda$CDM}
Again, we use the most basic $\Lambda$CDM model, optimizing the two
coefficients, $\alpha$ and $\beta$, from the L-$\sigma_v$ correlation, with $H_0$
and $\Omega_{\rm m}$ as the free cosmological parameters in this implementation. The
joint best fit for the \hii~and cosmic chronometer data sets is shown
in Figures~\ref{fig.10} and \ref{fig.11}, respectively. The optimized
parameter values corresponding to the optimized fit are presented in
Table~\ref{table4}, with the $1-2\sigma$ confidence regions shown in
Figure~\ref{fig.12}.

\subsection{The $R_{\rm h}=ct$ universe}
The $R_{\rm h}=ct$ universe has only one free cosmological parameter, $H_{0}$,
apart from the two nuisance parameters characterizing the luminosity 
of the HII galaxies, $\alpha$, and $\beta$, that must be optimized simultaneously
with the Hubble constant. The best fit for this
cosmology is shown in Figures~\ref{fig.13} and \ref{fig.14}, while the optimized
parameters are presented in Table~\ref{table4} and the $1-2\sigma$ confidence
regions are shown in Figure~\ref{fig.15}.

\section{Discussion}
\begin{figure}
\centerline{\includegraphics[angle=0,scale=0.63]{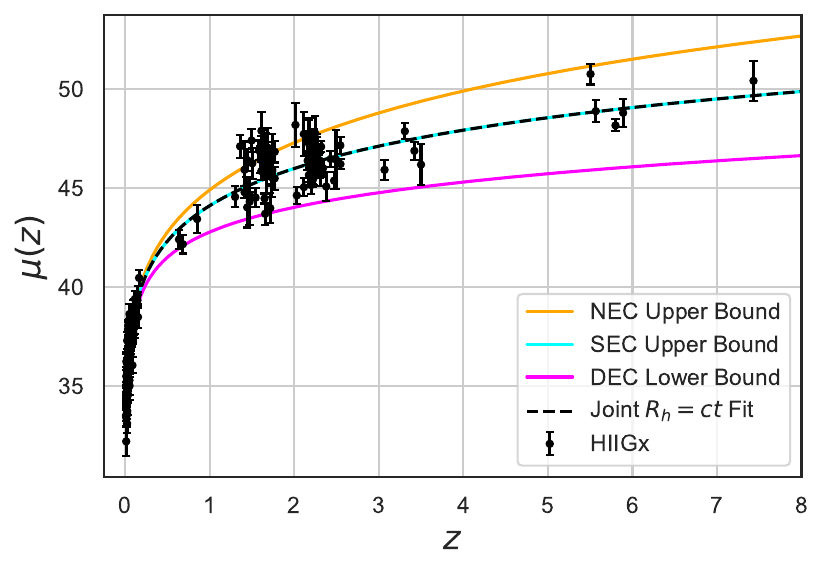}}
\caption{Same as Figure~\ref{fig.10}, except now for $R_{\rm h}=ct$. In
this cosmology, the SEC is identical to the model's predicted
$\mu(z)$, so all of the energy conditions are completely
satisfied.}\label{fig.13}
\end{figure}

\begin{figure}
\centerline{\includegraphics[angle=0,scale=0.63]{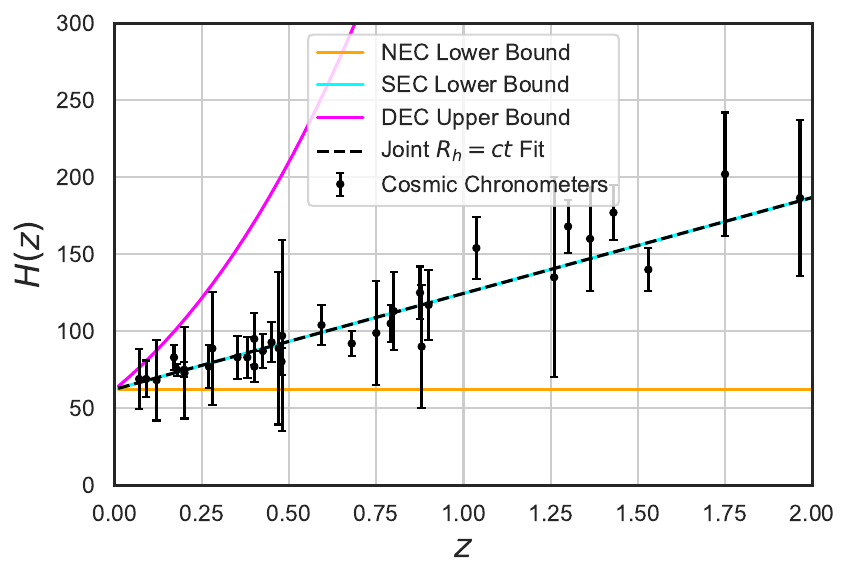}}
\caption{Same as Figure~\ref{fig.11}, except now for $R_{\rm h}=ct$. In
this cosmology, the optimized fit to the cosmic chronometer data is fully
consistent with all of the energy conditions.}\label{fig.14}
\end{figure}

\begin{figure}
\centerline{\includegraphics[angle=0,scale=0.47]{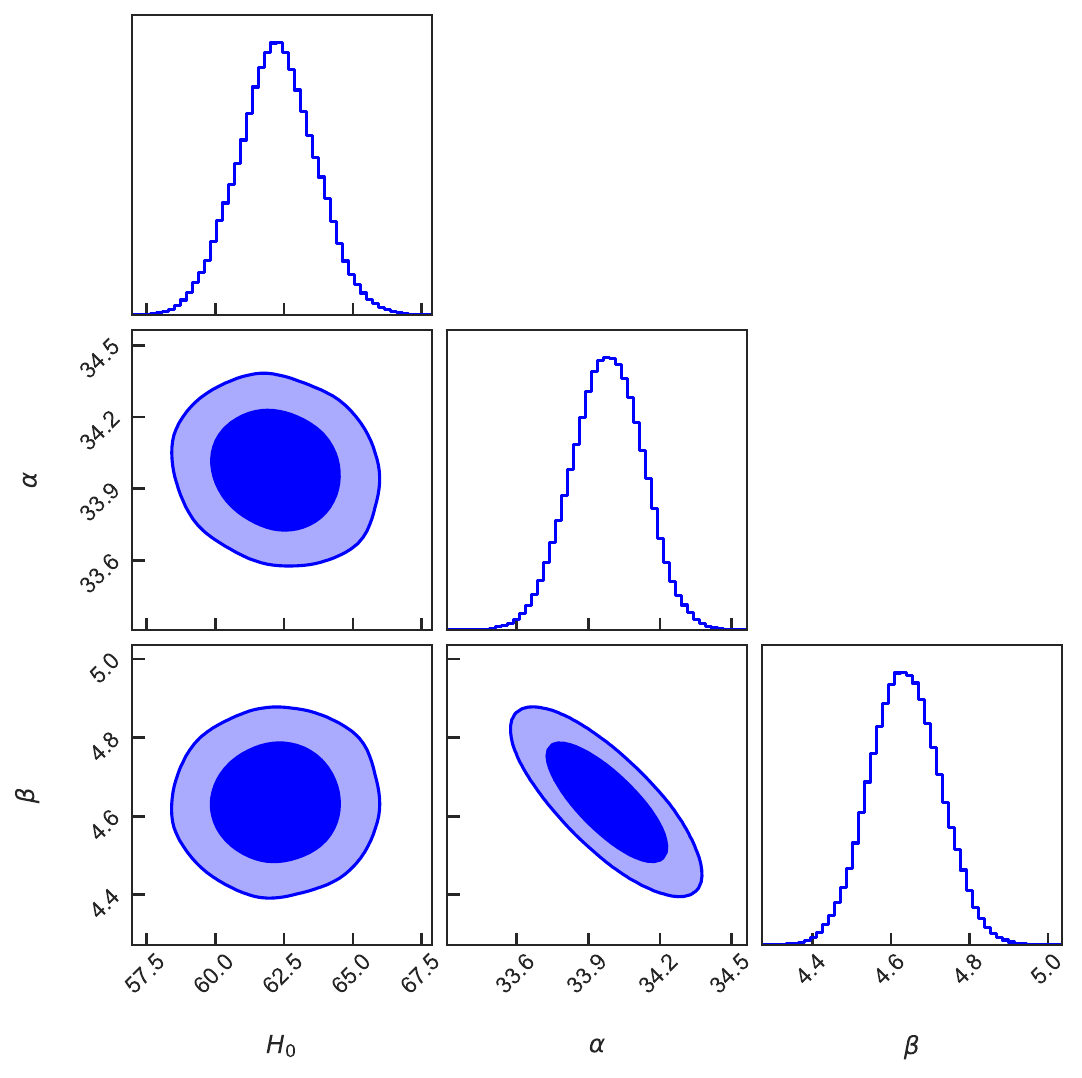}}
\caption{1D probability distributions and 2D regions with the $1-2\sigma$
contours corresponding to the parameters $\alpha$, $\beta$, and $H_0$ in the
$R_{\rm h}=ct$ universe fit (figs.~\ref{fig.13} and \ref{fig.14}) to the
combined \hii~galaxy $+$ cosmic chronometer data.}\label{fig.15}
\end{figure}

The weak energy condition and the null energy condition imply 
timelike and null observers always see a non-negative energy density. A violation 
of any of these would allow the formation of infinite amounts of matter in finite 
regions of spacetime \cite{HawkingEllis:1973}. The dominant energy condition 
imposes the weak energy condition and causality of energy and momentum fluxes. A 
violation of the DEC would imply transfer of energy at speeds greater than that of 
light. Both, $\Lambda$CDM and $R_{\rm h}=ct$ satisfy these three ECs,
based on the \hii~galaxy and cosmic chronometer data we have analyzed in this
paper (see figures~\ref{fig.1}, \ref{fig.4}, \ref{fig.6}, \ref{fig.8}, \ref{fig.10}, 
\ref{fig.11}, \ref{fig.13}, and \ref{fig.14}.) The same is not true of the strong
energy condition, however. This constraint requires that matter gravitates towards 
matter, i.e., that gravity is always attractive. In an FLRW universe, this implies 
that the cosmic expansion must never be accelerating, as stated mathematically in
Equation~(\ref{SEC}). The standard model violates this condition on 
two occasions. The first time during the postulated period of inflation 
\cite{2023AnP...53500157M}; the second during its predicted accelerated expansion 
of the Universe due to the dominance of a cosmological constant at low redshifts.
This second instance of the SEC violation is manifested in Figures~\ref{fig.1}, 
\ref{fig.2}, \ref{fig.6}, \ref{fig.10} and \ref{fig.11}, which show that the
best fit flat-$\Lambda$CDM cosmology requires the effects of antigravity to
account for the data, regardless of whether we consider the \hii~galaxies
and cosmic chronometers on their own, or whether we carry out a joint analysis. 
In sharp contrast, not only is the $R_{\rm h}=ct$ universe completely consistent
with all four of the energy conditions, but it is also strongly favoured
over flat-$\Lambda$CDM by the low-$z$ measurements.

The Bayes Information Criterion (BIC) \cite{1978AnSta...6..461S},
\begin{equation} \label{BIC_Definition}
BIC = -2\ln\mathcal{L} + f\ln(n)\,,
\end{equation}
where $\ln \mathcal{L}$ is the maximized log likelihood, $f$ is the number
of free parameters, and $n$ is the number of data points, is an asymptotic 
approximation to the outcome of Bayesian Inference (as $n \rightarrow \infty$). 
For the joint analysis, the BIC yields a probability of $\sim 92\%$
versus only $\sim 8\%$ for $R_{\rm h}=ct$ and flat-$\Lambda$CDM, respectively.
The $\Delta$BIC $\approx 5$ for this model comparison suggests that the
evidence in favour of the former model is therefore `quite strong'
(see ref.~\cite{2013MNRAS.432.2669M} for a more detailed description of
the BIC and other information criteria).

As we pointed out earlier, the Hubble parameter is cosmology-dependent. 
Its values determined by {\it Planck} (i.e., $=67.4 \pm 0.5 \: \rm km\,s^{-1}\,Mpc^{-1}$) 
\cite{2020A&A...641A...6P}, and that from Type Ia Supernovae ($=73.04 \pm 1.04 \: 
\rm km\,s^{-1}\,Mpc^{-1}$) \cite{2022ApJ...934L...7R} are based on the use of
$\Lambda$CDM as the background cosmology. The value of $H_0$ optimized
in $R_{\rm h}=ct$ is smaller (see, e.g., Table~4) than that in $\Lambda$CDM,
but fully consistent with all previously optimized values using other
data \cite{Melia:2026}.

A second possible concern with this work is that the analysis of the \hii~galaxies 
and cosmic chronometers in flat-$\Lambda$CDM yields somewhat incompatible values 
for $\Omega_{\rm m}$ when using the data sets separately. Such a disagreement 
suggests that the data acquisition may not have been ideal, possibly missing
some source of systematic uncertainty, or simply underestimating the overall error.
The fact that some data points in Figures~\ref{fig.13} and \ref{fig.14} lie above 
(and below) the SEC Bound points to this possible underestimation. These data,
however, have been used previously to test cosmological models and optimize
their parameters, and the results we have obtained in this work are consistent 
with those earlier measurements. Quite reasonably, all three of the approaches we
have taken for the analysis in this paper provide the same model selection outcome.
The low-$z$ data seem to favour the linear expansion predicted by
$R_{\rm h}=ct$, which is also consistent with all four energy conditions 
from general relativity, over an accelerated expansion that violates 
the strong energy condition. 

\section{Conclusion}
\hii~galaxies and cosmic chronometers have been used before for model
selection among these two cosmologies
\cite{2016MNRAS.463.1144W,2025MNRAS.542L..19W,2013MNRAS.432.2669M},
in each case producing an outcome favouring $R_{\rm h}=ct$. With
the improvement in the source catalogues, these comparisons now
favour this model even more robustly, particularly when using
a joint analysis of the two data sets.

This in itself provides rather compelling evidence in favour of this
model over the current standard scenario, but in this paper, we have
also demonstrated that this comparison draws distinctly different
perspectives within the framework of the energy conditions in general
relativity. While $\Lambda$CDM violates at least one of these
constraints at low redshifts, $R_{\rm h}=ct$ is completely consistent
with all of them.

This outcome, however, should not be viewed in isolation. It adds
to the already large body of evidence favouring $R_{\rm h}=ct$
over $\Lambda$CDM. And very interestingly, the inconsistency
between the standard model and the energy conditions arises in
several areas, most noticeably during the hypothesized
inflationary expansion shortly after the Big Bang. Ironically,
the supposed existence of the inflaton field is often touted
as `evidence' against the viability of the classical energy
conditions. Yet the observations today are suggesting more
and more that inflation is inconsistent with all of the
available data, particularly the anisotropies in the CMB
temperature field \cite{2013PhLB..723..261I,Liu:2020}.

We suggest that one should therefore view the violation of
the energy conditions by inflation as a compelling argument
against its viability, a point that has been made more formally
in ref.~\cite{2023AnP...53500157M}. Of course, inflation is not necessary
in $R_{\rm h}=ct$, which has no horizon problems and the fluctuations
producing the large scale structure emerged naturally at the Planck
scale \cite{2013A&A...553A..76M,2018EPJC...78..739M}.

The success of this analysis suggests that one ought to pay more
attention to the energy conditions, not only in the very early
Universe, but in the context of all cosmological observations,
at both high and low redshifts.

\vskip 0.2in
\noindent{\bf Data Availability Statement:} This manuscript does not
have any associated data.

\vskip 0.2in
\noindent{\bf Code Availability Statement:} This manuscript does not
have any associated code/software.

{\acknowledgement
We are grateful to the anonymous referee for helpful comments.
\endacknowledgement}

\bibliographystyle{JHEP3}
\bibliography{references}

\end{document}